\documentclass[submission,copyright,creativecommons]{eptcs}
\providecommand{\event}{GandALF 2015}
\usepackage[english]{babel}
\usepackage[latin1]{inputenc}
\usepackage{stmaryrd}
\usepackage{hyperref}
\usepackage{epsf}
\usepackage{epsfig}
\usepackage{graphicx}
\usepackage{gastex} 
\usepackage{amssymb}
\usepackage{amsfonts}
\usepackage{amsmath}
\usepackage{multirow}
\usepackage{latexsym}
\usepackage{verbatim}
\usepackage{xcolor}
\usepackage{color}
\usepackage{multirow}
\usepackage{algorithmic}
\usepackage{algorithm}
\usepackage{listings}

\usepackage{xparse}
\NewDocumentEnvironment{ctbl}{mmm} 
    { 
        \begin{table}[htbp]
        \begin{centering}
        \begin{tabular}{#1}
    }
    { 
        \end{tabular}
        \caption{#2}
        \label{#3}
        \end{centering}
        \end{table}
    }
\NewDocumentEnvironment{cfig}{mm} 
    { 
        \begin{figure}[htbp]
        \begin{centering}
    }
    { 
        \caption{#1}
        \label{#2}
        \end{centering}
        \end{figure}
    }

\newcommand{\tuple}[1]{\langle #1 \rangle}

\newsavebox{\sembox}
\newlength{\semwidth}
\newlength{\boxwidth}

\lstdefinelanguage{math-pseudocode}{
    morekeywords={
        Procedure,
        Begin,End,
        Input,
        Output,
        Var,
        choose,where,
        until,while,
        do, od,
        if, then, else, elseif, fi,
        return
    },
    sensitive=false,
    morecomment=[l]{//},
    morecomment=[s]{/*}{*/},
    morestring=[b]",
    mathescape=true,
}
\lstnewenvironment{mathpseudocode}[1][]
    {\lstset{language=math-pseudocode,float=ht,#1}}
    {}
\lstnewenvironment{mathinlinepseudocode}[1][]
    {\lstset{language=math-pseudocode,frame=none,#1}}
    {}

\floatname{algorithm}{Algorithm}

\newcommand{\mathactivatecomma}{%
  \begingroup\lccode`~=`\,
  \lowercase{\endgroup\edef~}{\mathchar\the\mathcode`\,\penalty0 }}

\newcommand{\erase}[1]{}

\newcommand{\mincl}{\sqsubseteq}
\newcommand{\munion}{\oplus}
\newcommand{\Nat}{{\mathbb{N}}}
\newcommand{\multisets}[1]{{#1}^M}
\newcommand{\mdiff}{\ominus}

\newtheorem{theorem}{Theorem}[section]

\newtheorem{proposition}[theorem]{Proposition}

\newenvironment{proof}[1][Proof]{\begin{trivlist}
\item[\hskip \labelsep {\bfseries #1}]}{\end{trivlist}}
\newenvironment{definition}[1][Definition]{\begin{trivlist}
\item[\hskip \labelsep {\bfseries #1}]}{\end{trivlist}}

\newcommand{\qed}{\nobreak \ifvmode \relax \else
      \ifdim\lastskip<1.5em \hskip-\lastskip
      \hskip1.5em plus0em minus0.5em \fi \nobreak
      \vrule height0.75em width0.5em depth0.25em\fi}

\def\titlerunning{Wsts with history}
\def\authorrunning{Abdulla-Delzanno-Montali}

\title{Well Structured Transition Systems with History}
\author{Parosh A. Abdulla 
\institute{Uppsala University}
\and 
Giorgio Delzanno
\institute{University of Genova}
\and 
Marco Montali\textbf{}
\institute{Free University Bolzano}
}

\begin{document}
\maketitle
%
%
\begin{abstract}
We propose a formal model of concurrent systems in which the history of a computation is explicitly represented as a collection of events that provide a view of a sequence of configurations. 
In our model events generated by transitions become part of the system configurations leading to operational semantics with historical data. This model allows us to formalize what is usually done in symbolic verification algorithms. 
Indeed, search algorithms often use meta-information, e.g., names of fired transitions, selected processes, etc., to reconstruct (error) traces from symbolic state exploration. The other interesting point of the proposed model is related to a possible new application of the theory of well-structured transition systems (wsts).
In our setting wsts theory can be applied to formally extend the class of properties that can be verified using coverability to take into consideration (ordered and unordered) historical data.
This can be done by using different types of representation of collections of events and by 
combining them  with wsts by using closure properties of well-quasi orderings.
\end{abstract}

\section{Introduction}
Well-structured transition systems (wsts) are an important class of infinite-state systems for which it is possible to 
decide algorithmically verification problems like coverability and boundedness.
This class of systems include models like Lossy Channel Systems, Petri Nets, Datanets, Multiset rewriting with Constraints, and 
Timed Networks \cite{ACJT96,AJ01,AN00,ADV11,DR13,DelzannoMSR,FS01,LNORW08}. 
The theory behind wsts is based on two key points: (a) a well-quasi ordering is introduced to compare configurations w.r.t. their information contents, (b) transitions are required to be monotone with respect to the considered ordering.
The combination of these two properties lead to a general framework in which it is possible to algorithmically 
decide a class of reachability problems defined by considering target states larger than a given configuration.
The decision procedure is based on symbolic state exploration. Symbolic representations are based on the finite-basis property of well-quasi ordering, namely every upward closed set can be finitely generated.
The minimal elements of an upward closed set are then used as symbolic representations of infinite-sets of 
configurations \cite{ACJT96,AJ01,FS01}.
Apart from models like Petri nets and Lossy Channel Systems, the theory of wsts has been applied to study computational models resulting from a combination of different types of systems like pushdown automata with well-quasi ordered locations/data \cite{AADP13,CO13,CO14}, asynchronous systems defined by extending pushdown systems with an external 
memory \cite{ChadhaV09}, and others.

In the present paper we use the theory of wsts as a tool to study properties of transition systems extended with history information. In this setting one possible formalization of the extended notion of transition systems 
is based on rules that generate events. In the operational semantics events generated during the application of transitions are collected in 
a read-only memory that acts as a sort of log.
The generated log can be queried in order to formalize properties related to the sequence of transitions that yield a given configuration. Events can be defined as simple labels or as structured data that can share information with configurations
(e.g. an event contains a piece of data generated by a transition).
By using this idea, it is possible to define a generalized version of the coverability problem that takes into consideration an ordering on states and an ordering on histories (logs).
We refer to the resulting coverability problem as History Coverability (HCOV).
HCOV can be instantiated in order to formulate properties like  provenance and correspondence.

In this paper we investigate this idea in two steps.
\begin{itemize}
\item
We first study the problem of preserving wsts properties when extending a transition system with events and 
histories/logs.  In this setting we apply general results on (combination of) well-quasi orderings like Highman's Lemma in order to define  conditions under which HCOV is still decidable when the underlying transition system is a wsts. 
To obtain positive results in a compositional way, it seems necessary to consider events that are independent from 
configurations. In this sense we can think about extended transition systems in which we plug an ad hoc memory in which to collect events that form a log of a given computation. 
\item
We then consider a more general notion of log in which states and events are no more independent, e.g., they can share common data or time-stamps used to enrich the logs collected during a computation.
In this settings it seems more difficult to obtain positive results by using a compositional approach
based on closure properties of well-quasi orderings.
For this reason, we propose a different approach:
\begin{itemize}
\item we first fix the structure underlying the considered systems, e.g., we consider configurations and logs as multisets of predicates/terms;
\item we then apply a general purpose language called MSR(Id), an instance of  multiset rewriting with constraints in which values are ordered identifiers,  as a meta-language in which to encode different types of transitions systems with history and logs. 
\end{itemize}
\end{itemize}
For the considered models, we exploit properties of the host formalism in order to give conditions under which it is possible to decide the HCOV problem even in presence of dependency relations between configurations and logs.
The resulting framework shows a potential new application of the theory of well-structured transition systems  to a class of properties like correspondence and provenance that go beyond coverability.
\section{Transition Systems}
Given a quasi order $\tuple{S,\leq}$, an upward closed set of states is a subset $U\subseteq S$
such that for any $s\in U$, if $s\leq s'$ then $s'\in U$.
Given a set $B$ we say that $B$ generates the upward closed set 
$B\uparrow=\{s|s'\in B,\ s'\leq s\}$.
\begin{definition}
A well quasi ordering $\tuple{S,\leq}$ is a quasi ordering such that for every infinite sequence
of elements $s_1 s_2\ldots$ there exist $i<j$ such that $s_i\leq s_j$.
A well quasi ordering has the finite basis property, i.e., every upward closed set $U\subseteq S$
is generated by a finite set $B$.
\end{definition}
Let $S$ be an infinite set of configurations.
A transition system $T$ is a tuple $T=\tuple{S,\rightarrow,s_0}$ such that 
$\rightarrow\subseteq S\times S$ is the transition relation, and $s_0$ is the initial state.
We use $s_1\rightarrow s_2$ to denote a pair $\tuple{s_1,s_2}\in\rightarrow$.
A computation is a sequence of states $s_0 s_1 s_2\ldots$ s.t. $s_i\rightarrow s_{i+1}$ 
for $i\geq 0$. 
Given a transition system $T$, the (one step) predecessor states of a set of configurations $A$ is defined as 
$Pre_T(A)=\{s|s\rightarrow t,\ t\in A\}$.
The whole set of predecessor states of a set of configuration $A$ is defined as 
$Pre_T^*(A)=\bigcup_{i\geq 0} Pre_T^i(A)$, where 
$Pre_T^0(A)=A$, and 
$Pre_T^{i+1}(A)=Pre_T(Pre_T^i(A))$ for $i\geq 0$.
We will often use $Pre(a)$ instead of $Pre_T$, when $T$ is clear from the context.

A transition system $T$ is monotone w.r.t. $\leq$ if for every $s_1,s_2,s_3$ s.t. 
$s_1\rightarrow s_2$ and $s_1\leq s_3$ there exists $s_4$ s.t. $s_3\rightarrow s_4$ and $s_2\leq s_4$.
In other words the diagram formed by $s_1,s_2,s_3,s_4$ combining $\rightarrow$ and $\leq$ commutes.
\begin{definition}
A transition system $T$ is well structured (wsts) if  $T$ is monotone w.r.t. a well quasi ordering $\leq$ on configurations.
\end{definition}
We need two additional properties to obtain positive results for verification problems.
\begin{definition}
A wsts$^*$ is a wsts that satisfies the  following additional conditions:
\begin{itemize}
\item Given a basis $B$ of an upward closed set of configurations $U$, it is possible
to algorithmically compute a basis $B'$ of the set of predecessor states $Pre_T(U)$ of $U$, 
\item It is possible to algorithmically check whether $s_0$ belongs or not to a set of an upward closed set of 
configurations.
\end{itemize}
\end{definition}
The {\em Coverability Problem} (COV) is defined as follows.
Given a transition system $\tuple{S,\rightarrow,s_0}$, a quasi order $\leq$ on $S$, 
and a state $s_1\in S$, we want to check whether or not there exists a state $s_2\in S$ 
and a computation from $s_0$ to $s_2$ s.t. $s_1\leq s_2$.
The problem can be generalized by considering an infinite set $I$ of initial configurations as follows.
Given a state $s_1\in S$, we want to check whether or not there exists an initial state $s_0\in I$, 
a state $s_2\in S$ and a computation from $s_0$ to $s_2$ s.t. $s_1\leq s_2$.

COV is decidable for wsts$^*$ transition systems \cite{ACJT96,AJ01,FS01}.
The algorithm that can be used to decide the problem is based on symbolic backward reachability.
Specifically, let $B=\{s_1\}$ be the basis that generates the upward closed set $B\uparrow$, 
i.e., the infinite set of configurations generated by taking all states that are larger, w.r.t. $\leq$,
than $s_1$, namely $B=\{s| s_1\leq s\}$.
Symbolic backward reachability computes the chain (w.r.t. subset inclusion) of sets defined as
\begin{itemize}
\item $I_0=B$, 
\item $I_{i+1}=I_i\cup Pre(I_i)$ for $i\geq 0$.
\end{itemize}
Clearly $I_i\subseteq I_j$ for $i\leq j$.
Furthermore, it can be shown that the chain stabilizes (i.e. it reaches a least fixpoint) 
if $\leq$ is a wqo. 
Namely, if $\leq$ is a wqo, then there exists $k$ s.t. $I_{k+1}\uparrow=I_k\uparrow$.
When the algorithm has reached a least fixpoint as step $k$, $I_k$ is a finite basis
for $Pre^*(B)$, i.e., $Pre^*(B)=I_k\uparrow$.
To test COV we just need to check whether $s_0\in I_k\uparrow$ a decidable test by definition of wsts.
The above described (ideal) algorithm can be implemented using different types of heuristics.
For instance, we can apply a subsumption test to discard elements of $Pre(I_i)$ that are redundant 
w.r.t. information that is already present in $I_i$.

Constraints or other forms of symbolic representations of upward closed sets of configurations can be applied
to lift the algorithm to procedures that combine external solvers or decision procedures.
For instance, when considering multisets defined over a finite set of symbols with multiset inclusion,
we can use numerical inequalities the form $X_s\geq c$ to keep track of upper bounds on the number of 
occurrences of instances of symbol $c$ (i.e. at least $c$ occurrences).
This representation can then be used to apply numerical solvers to handle upward closed sets of configurations.

\section{Transition Systems with History}
In this section we defined an extended notion of transition systems with an explicit representation of events
generated during a computation. Events can be simple letters (as customary when reasoning on languages generated by 
transition systems) or work as a sort of external memory in which to store not only event labels but pieces of data occurring in a configuration. In this paper we focus our attention on logs defined via a read-only memory
and consider conditions under which it is possible to extended positive properties of wsts 
to transition systems with logs.

Let $S$ be an infinite set of configurations and $E$ be an infinite set of events.
Furthermore, we say that $H$ is a set of histories of $E$ 
if $H$ is an infinite set with: (a) an element ${\mathbf 0}\in H$, and (2) a binary operation 
$+:E\times H\rightarrow H$. 

For a transition system $\tuple{S,\rightarrow,s_0}$ and a set of events $E$,
a transition system with history is a tuple $\tuple{S,E,\rightarrow_h,s_0}$ such that 
for $\rightarrow_h\subseteq S\times S\times E$, the transition relation with history, 
it holds that for each $s\rightarrow_h s'[e]$ there exists a transition 
$s\rightarrow s'$ (i.e. the projection of $\rightarrow_h$ on $S\times S$ is 
$\rightarrow$). $s_0$ is the initial state.

A configuration with history is a pair $\tuple{s,h}$, written $s[h]$, s.t. $s\in S$ and $h\in H$ where $H$ 
is the set of all possible histories with elements in $E$.
We now define the notion of wsts with history. 
For this purpose, we need to introduce an ordering $\sqsubseteq$ between histories (logs).
\begin{definition}
A wsts with history (hwsts) is a tuple $\tuple{S,E,\rightarrow_h,s_0,\leq,+,\sqsubseteq}$ 
such that 
\begin{itemize} 
\item $\tuple{S,\rightarrow,s_0}$ is a wsts,
\item $\rightarrow_h$ is a transition relation with history built on top of $S$ and $E$,
\item if $s\rightarrow_h s'[e]$, and $s\leq t$, then there exists $t\rightarrow_h t'[e']$ 
s.t. $s'\leq t'$ and $e\sqsubseteq e'$.
\item $+:E\times H\rightarrow H$ satisfies the following property 
if $h\sqsubseteq h'$ and $e\sqsubseteq e'$, then  $e+h\sqsubseteq e'+h'$ for any $e,e'\in E$;
\item $\tuple{H,\sqsubseteq}$ is a well-quasi ordering.
\end{itemize}
\end{definition}
A computation is a sequence of configurations with history $s_0[h_0] s_1[h_1] s_2[h_2]\ldots$ s.t. 
$h_0={\mathbf 0}$, $s_i\rightarrow_h s_{i+1}[e_i]$ and $h_{i+1}=e_i+h_{i}$ for $i\geq 0$.

We now introduce the decision problems, called {\em History Coverability Problem} (HCOV), 
we will focus our attention on in the rest of the paper.
\begin{definition}
Given a hwsts $\tuple{S,E,\rightarrow_h,s_0,\leq,+,\sqsubseteq}$, a state $s_1\in S$ and a history $h$, 
HCOV consists in checking whether there exists a computation from $s_0[{\mathbf 0}]$ 
that can reach a configuration with history $s'[h']$ s.t. $s_1\leq s'$ and $h\sqsubseteq h'$.
\end{definition}
\subsection{General Conditions for Decidability of HCOV}
In this section we apply the theory of well-structured transition systems to obtain general conditions
on the decidability of HCOV.
We first introduce an ordering on configurations with histories.
Namely, we define $s_1[h_1]\preceq s_2[h_2]$ if and only if $s_1\leq s_2$ and 
$h_1\sqsubseteq h_2$.
The following property then holds.
\begin{proposition}\label{precwqo}
The ordering $\preceq$ is a well quasi ordering.
\end{proposition}
\begin{proof}
For qo $\tuple{A_1,\leq_1}$ and $\tuple{A_2,\leq_2}$, consider the qo 
$\tuple{A_1\times A_2,\leq}$ such that $\tuple{a_1,a_2}\leq \tuple{a_1',a_2'}$
iff $a_i\leq_i a_i'$ for $i:1,2$. 
The generalized version of Highman's lemma  states that if $\leq_1$ and $\leq_2$ are wqo's, 
then the above defined ordering $\leq$ is still a wqo.
\\
We can apply the lemma to configurations of the form $s[h]$ with $s\in S$ and $h\in H$, assuming 
that both $\leq$ and $\sqsubseteq$ are wqo's.
\hfill\qed
\end{proof}
A hwsts satisfies then following property.
\begin{proposition}
A hwsts is monotone w.r.t. $\preceq$, i.e., 
if $s_1[h_1]\rightarrow_h s_2[h_2]$ and  $s_1[h_1]\preceq s_3[h_3]$, 
then there exists 
 $s_3[h_3]\rightarrow_h s_4[h_4]$ s.t.  $s_2[h_2]\preceq s_4[h_4]$.
\end{proposition}
\begin{proof}
By definition, 
$s_1[h_1]\rightarrow_h s_2[h_2]$ implies that there exists  $s_1\rightarrow_h s_2[e]$ s.t. $h_2=e+h_1$.
By definition of hwsts, 
if $s_1\leq s_3$, then there exists  $s_3\rightarrow_h s_4[e']$ s.t. $s_2\leq s_4$ and $e\sqsubseteq e'$.
\\
By definition of $+$ and since $e\sqsubseteq e'$, we have that $e+h_1\sqsubseteq e'+h_3$, 
hence $h_2\sqsubseteq h_4$, since $h_2 = e + h_1$ and $h_4 = e' + h_3$.
\hfill\qed
\end{proof}
We now have a wsts transition system in which we can represent the history of a computation by composing 
events to form histories.  
It remains to define conditions under which we can algorithmically compute predecessor states.
Now consider a wqo $\preceq$ associated to an hwsts.
Every upward closed sets $A$ of configurations with history can be represented by a finite basis, 
i.e., a finite set of configurations and histories.
Let us now call hwsts$^*$ a hwsts such that for any upward closed sets of configurations with history represented by a finite basis $B$, we can algorithmically compute a finite basis $B'$ for  $Pre(B)$. 
The following property then holds.
\begin{proposition}\label{proppre}
Fix an hwsts$^*$ and a basis $B$ of an upward closed set of configurations with history, 
we can algorithmically compute a finite representation of $Pre^*(B)$. 
\end{proposition}
\begin{proof}
Starting from $B$, we can iterate the application of $Pre$ and compute finite basis of intermediate results.
The wqo condition ensures termination.
\hfill\qed
\end{proof}
The previous property can be exploited in order to define decision procedures for history dependent  properties.
From Prop. \ref{proppre}, it follows that HCOV is decidable for hwsts$^*$.
The algorithm is based on a saturation procedure that computes a finite representation of $Pre^*(B)$
where $B$ is the basis of an upward closed set defined by state $s_1[h]$.
To give examples of history structures that satisfy the conditions of our results, 
we have to instantiate $H$ and $+$.
As an example, consider a domain $H$ defined as the set of multisets of events in $E$ and $+$ 
as the multiset constructor, i.e., $e+h=\{e\}\oplus h$, where $\oplus$ denotes multiset union.
Let us assume that $\tuple{H,\subseteq}$ is a well quasi ordering w.r.t. sub-multiset inclusion
(e.g. $E$ is a finite set and $\sqsubseteq$ is equality over elements).
Then, we can apply the decision procedure of Prop. \ref{proppre} to decide 
coverability for state $s$ along path that contain a given multiset of events $e_1,\ldots,e_n$.

Now consider a domain $H$ defined as the set of words in $E^*$ where $+$ 
is just concatenation, i.e., $e+h=e.h$.
Let us assume that $\tuple{H,\subseteq}$ is a better quasi ordering w.r.t. subword inclusion
(again $E$ is a finite set and $\sqsubseteq$ is equality over elements).
Then, we can apply the decision procedure of Prop. \ref{proppre} to decide 
coverability for state $s$ along path that contain a given sequence of events $e_1,\ldots,e_n$.
\subsection{Automata with History}
The first example that we consider is an extension of finite-state automata with history.

A finite-state automaton, interpreted as a computational model and not as a language acceptor, 
is a tuple $A=\tuple{Q,\delta,s_0}$ where $Q$ is a finite set of states, $\delta$ 
is a transitions relation $\delta\subseteq Q\times Q$ and $s_0\in Q$.
An execution is a sequence of states $s_0 s_1 s_2\ldots$ s.t. $\tuple{s_i,s_{i+1}}\in\delta$ for $i\geq 0$.
Given states $s_0$ and $s_1$, the reachability problem consists in checking whether there exists a computation 
from $s_0$ to $s_1$. Let us now extend finite-state automata in order to maintain history information.
We use $t=s\rightarrow s'$ to denote a single transition $\tuple{s,s'}\in\delta$.
Let us now consider the standard way to associate words to computations based on labeled transitions.
In our setting labels can be viewed as events added to the current log as in the transition $s\rightarrow s'[e]$.
The semantics is defined by collecting events in the current history.
Namely, for $t=s\rightarrow s'[e]$,  $s'[e.h]$ is a successor of $s[h]$ in which the history $h$ 
is extended with event $e$. 
 
We now reformulate HCOV in this setting.
Given states $s_0$ and $s_1$ and events $e_1$ and $e_2$, 
we are interested in checking whether there exists an history $h$ and 
a computation from $s_0[\epsilon]$ to $s_1[h]$ such that $e_1 e_2$ is a subword of $h$.
If $e_1$ and $e_2$ are associated to transitions $t_1$ and $t_2$, 
this amounts to check whether there exists a computation in which $t_2$ can be fired after $t_1$. 

When events are elements from a finite alphabet, histories correspond to words generated by an automaton.
HCOV can then be solved using language inclusion by comparing the language generated by automaton $A$ 
with a regular language that encodes sequences of events we are interested in.
We observe that since words are wqo w.r.t. subword relation, from Prop. \ref{proppre}, 
we have that HCOV can be solved via the backward reachability algorithm  that, from a finite basis of the form $s[w]$ where $s\in Q$ and $w$ is an history. computes all predecessor states.
This property still holds for logs defined by different data structures, e.g., when replacing words with counters 
that keep track of the number of occurrences of events in a computation (a sort of Parikh image).
In this setting we consider a finite number of constants $e_1,\ldots,e_n$ that represent occurrences of events
The semantics is defined by collecting events in a multiset instead of a word, i.e.,
Namely, for $t=s\rightarrow s'[e]$,  $s'[e\oplus h]$ is a successor of $s[h]$, 
where $e\oplus h$ denotes the multiset obtained by adding $e$ to multiset $h$. 

The resulting transition system is monotone w.r.t. equality over states and multiset inclusion over logs.
Logs can be viewed as counters that grow monotonically and count the number of occurrences of events in a computation.
We can now use $s[\varphi]$ where $\varphi$ is a constraint over the counters of the form 
$c_1\geq a_1,\ldots,c_n\geq a_n$ with $a_1,\ldots,a_n$ natural numbers,  to obtain a class of queries for which HCOV is decidable.
\subsection{Petri Nets with History}
The second example that we consider is related to Petri Nets with history.

A Petri net is a tuple $\tuple{P,T,M_0}$ where $P$ is a finite set of places, $T$ is a finite set of transitions i.e. 
a subset of  $P\times P$, and $M_0$ is the initial marking. 
A marking is mapping $M:P\rightarrow\Nat$ that associates a $M(p)$ tokens to a given places.
Tokens can be viewed as indistinguishable process instances (i.e. processes without identifiers or internal data).
Places can be viewed as process states, i.e., a token in place $p$ corresponds to a process in state $p$.
A marking can be viewed then as an abstract representation of a global configuration of a concurrent system.
Since the number of places is finite a marking $M$ can be viewed as a  vector of natural numbers $\tuple{c_1,\ldots,c_n}$
where $c_i$ is the number of tokens in place $p_i$ for $i:1,\ldots,|P|$ or as a multiset over $P$ such that the number 
of occurrences of symbol $p$ in $M$ corresponds to $M(p)$.

A transition $t$ describes a possible concurrent update of a finite number of tokens.
More formally, let $\multisets{P}$ be the class of multisets over $P$.
Assume let $t=\tuple{Pre,Post}$ with $Pre,Post\in P^\oplus$.
$t$ is enabled at marking $M$ if $Pre\mincl M$ using the multiset notation for markings ($\mincl$ is multiset inclusion).
If $t$ is enabled in $M$, the firing of $t$ yields  a new marking $M'$ defined as 
$M'=(M{\mdiff} Pre)\munion Post$ using the multiset notation for markings.
Namely, the tokens in $Pre$ are removed from $M$ and those in $Post$ are added to the resulting multiset.
An execution is a sequence of markings $M_0 M_1 M_2\ldots$ s.t. $M_{i+1}$ is obtained from $M_i$ by firing a transition for $i\geq 0$.
We use $M_0\triangleright M$ to denote an execution from $M_0$ to $M$, i.e., $M$ is reachable from $M_0$.
Given markings $M_0$ and $M_1$, the coverability problem consists in checking whether there exists a marking $M_2$ s.t. 
$M_0\triangleright M_2$ s.t. $M_1\mincl M_2$.
The coverability problems requires then to find a reachable marking that contains in each place at least as many tokens 
as those contained in $M_1$. This problem can be used to encode reachability of configurations that violate a safety property (e.g. a configuration in which a token is in an error place). 

Let us now extend Petri Nets in order to maintain history information.
We now consider histories defined via sequences of transition names $t_1t_2\ldots$
and transitions that emit events of the form $h_t$:
$$
\begin{array}{l}
Pre\rightarrow Post[h_t]
\end{array}
$$
The semantics with history is defined by collecting events in the current history.
Namely, for $t=\tuple{Pre,Post}$,  $(Pre\oplus M)[h]\triangleright (Post\oplus M)[h_t.h]$
denotes the extension of history $h$ with the event $h_t$.  
Since multisets are wqo w.r.t. the submultiset relation and words are wqo w.r.t. the subword relation, 
from Prop. \ref{proppre}, we have that HCOV is decidable via a backward reachability algorithm  
that works over finite basis of the form $M[w]$ where $M$ is a marking and $w$ is a history. 
\begin{theorem}
HCOV is decidable for Petri Nets with history.
\end{theorem}
\section{Transition Systems with History}
In the previous section we have defined separate conditions on states and histories to deduce wsts properties
on transitions that generate events collected during a computation.
This kind of reasoning can be applied to histories defined by elements that are independent from states,
e.g., symbols that represent events in the execution. However, there are situations in which it could be more convenient to maintain relations  between elements in the state and elements in the history.
Generalizing the notion of history transition system in order to maintain well-structuredness 
is not immediate without more information about the structure of configurations and events. In this section we consider a possible formulation of the considered properties within MSR(${\cal C}$)
a formal model of concurrent computation that combines rewriting and constraints.
The idea here is to exploit the expressiveness of the considered framework as a possible host language
in which to represent transition systems with history.
We will introduce MSR(${\cal C}$) in the following section.
\subsection{MSR(${\cal C}$)}
MSR(${\cal C}$) is a formal model for concurrent systems based on a combination of rewriting 
and constraints. A constraint system ${\cal C}$ is defined by formulas with free variables in $V$, 
an interpretation domain ${\cal D}$, and a satisfiability relation $\models$
for formulas in ${\cal C}$ interpreted over ${\cal D}$.
We use ${\cal D}\models_\sigma\varphi$ to denote satisfiability of $\varphi$ via 
a substitution $\sigma:Var(\varphi)\rightarrow {\cal D}$, where 
$Var(\varphi)$ is the set of free variables in $\varphi$.

For a fixed set of predicates $P$, an atomic formula with variables 
has the form $p(x_1,\ldots,x_n)$ where $p\in P$ and $x_1,\ldots,x_n\in V$.
A rewriting rule has the form $M\rightarrow M':\varphi$, 
where $M$ and $M'$ are multiset of atomic formulas with variables over $P$ and $V$, 
and $\varphi$ is a constraint formula over variables $Var(M\oplus M')$ 
occurring in $M\oplus M'$. We use $M=A_1,\ldots,A_n$ to denote a multiset of atoms.

MSR(Id) is the instance obtained by considering the constraint system $Id$ defined as follows.
\begin{itemize}
\item
Constraint formulas are defined by the grammar $\varphi::=\varphi_1,\varphi_2|x=y|x<y$ 
for variables $x,y\in V$.
Here $\varphi_1,\varphi_2$ denotes a conjunction of formulas $\varphi_1$ and $\varphi_2$.
\item 
The interpretation domain is defined over an infinite and ordered set of identifiers 
$\tuple{Id,=,<}$. 
\item For substitution $\sigma:V\rightarrow Id$, 
$x=y$ is interpreted as $\sigma(x)=\sigma(y)$, $x<y$ is interpreted as $\sigma(x)<\sigma(y)$, and
$\varphi_1,\varphi_2$ is interpreted as $\sigma(\varphi_1)\wedge\sigma(\varphi_2)$.
\end{itemize}
A constraint $\varphi$ is satisfied by a substitution $\sigma$ if $\sigma(\varphi)$ 
evaluates to $true$. An instance $M\sigma\rightarrow M'\sigma$ of a rule $M\rightarrow M':\varphi$ 
is defined by taking a substitution $\sigma:Var(M\oplus M')\rightarrow Id$ 
such that $\sigma(\varphi)$ is satisfied in the interpretation $Id$.

As an example, consider the rule $p(x,y),q(x)\rightarrow p(x,y),q(x),q(u):x<u$.
The intuition is that processes $p(x,y)$ and $q(z)$ synchronize when $x=z$ and 
generate a new instance $q(u)$ with $x<u$.
By associating natural numbers to identifiers,  
$p(1,2),q(1)\rightarrow p(1,2),q(1),q(4)$ and $p(3,10),q(3)\rightarrow p(3,10),q(3),q(8)$
are two instances of the considered rule.
We use $Inst(\Delta)$ to indicate the infinite set of instances of a set $\Delta$ of MSR rules.

A configuration is a multiset $N$ of atoms of the form $p(d_1,\ldots,d_n)$ 
with $d_i\in Id$ for $i:1,\ldots,n$. For a set $\Delta$ of rules and a configuration $N$, 
a rewriting step is defined by the relation $\triangleright$ s.t. 
$$N=(M\oplus Q)\triangleright (M'\oplus Q)=N'$$
for $(M\rightarrow M')\in Inst(\Delta)$.
A computation is a sequence of configurations $N_1\ldots N_m\ldots$ s.t. 
$N_i\triangleright N_{i+1}$ for $i\geq 0$.

The coverability problem for MSR(Id), MSRCOV, is defined as follows.
Given a specification $R$, an initial configuration $M_0$, and predicate $ok$,
is there a computation from $s_0$ to a configuration $M_1$ that contains at least one occurrence of predicate $ok$?

Coverability is undecidable in general, but decidable for monadic predicates only \cite{DelzannoMSR}.
In this setting we admit only predicates of the form $p(x)$ where $x$ is a variable that may occur in a constraint. 
MSR(Id) with monadic predicates subsumes Petri Nets and it has the same expressive power as 
Data Nets \cite{ADV11}.
It is important to observe that in a rule $M\rightarrow M':\varphi$ it is not required that all variables
occurring in $M'$ occur in $M$. A variable that occurs only in $M'$ can be instantiated
with an arbitrary identifier as variable $u$ in the above discussed rule 
$p(x,y),q(x)\rightarrow p(x,y),q(x),q(u):x<u$. Even for fixed instantiations of $x,y$ we can still
consider an infinite set of instances for variable $u$ (all values larger than the instantiation of $x$). 

The decision procedure for monadic MSR(Id) is based a symbolic representation of upward closed sets of configurations obtained as follows.
We consider constrained configurations of the form $\Psi=(p_1(x_1),\ldots,p_n(x_n):\varphi)$, 
where $\varphi$ is a constraint with variables in $x_1,\ldots,x_n$.
We then assign the following denotation to a constrained atom $\Psi$:
$$
Inst(\Psi)=\{M'\sigma\oplus Q|\Psi=(M:\varphi),\ \sigma:Var(M)\rightarrow Id,\ \sigma(\varphi)\ is\ satisfied\}
$$
Notice that in the denotation of $\Psi$ we consider all possible instances $M'$ of multiset $M$ 
as well as all possible configurations larger than $M$, i.e., that contain more processes.
\subsection{MSR(Id) as a Metalanguage for History Transition Systems}
We now show that MSR(Id) can be used as a meta-language to represent transition systems with history.
This allows us to infer good properties for transitions systems in which events and configurations
share common information (are in some relation). In particular, if the encoding of the transition system yields a specification in MSR(Id) with monadic predicates
only, then from decidability of MSRCOV we obtain decidability of HCOV.
\subsubsection{Petri Nets with history}
Let us go back to Petri Nets with history and consider transitions that emit events of the form $h_t$
(name of transitions), e.g., $Pre\rightarrow Post[h_t]$
The semantics with history is defined by collecting events in the current history.
Namely, for $t=\tuple{Pre,Post}$,  $Pre\oplus M[h]\triangleright Post\oplus M[h_t.h]$
denotes the extension of history $h$ with event $h_t$.  

The extended notion of history can be encoded in MSR by using timestamps as described next.
We first introduce a predicate $time(t)$ to associate a time stamp to each firing step.
Transitions with history are represented then as rewriting rules of the following form:
$$
\begin{array}{l}
Pre,time(t) \rightarrow Post,time(t'),h_{t}(t): t'>t
\end{array}
$$
We use predicate $h_{t}$ to denote an application of transition $t$.
A configuration in the resulting model consists of a marking $M$, a predicate 
$time(t)$, and a multiset of events $Ev$. 
\\
By construction, we have that if 
$(M_0\oplus T_0\oplus Ev_0) (M_1\oplus T_1\oplus Ev_1) \ldots (M_n\oplus T_n\oplus Ev_n)$, 
then $T_i=\{time(t_i)\}$ for  $i:1,\ldots,n$ and $t_1<t_2<\ldots t_n$.

The $time$ predicate can then be exploited in order to define queries on the history of a computation.
For instance, we can define an MSR rule of the form $h_{t_1}(x),h_{t_2}(y)\rightarrow ok: x<y$.
in order to check whether a given sequence of transitions, e.g., $t_1$ before $t_2$, can be fired during a 
computation.  Indeed, coverability w.r.t. to the initial configuration $M_0,time(t_0)$ and predicate $ok$ amounts to check whether there exists an execution that can reach a configuration in which $h_{t_1}(s),h_{t_2}(p)$ occur for $s<p$.
\subsubsection{Processes with data}
Consider now a multiset rewriting system with monadic predicates used as a model 
of processes with data.
Take for instance, the following rule:
$$
p_1(t_1),\ldots,p_n(t_n)\rightarrow q_1(s_1),\ldots,q_m(s_m)
$$
in which $t_1,\ldots,t_n,s_1,\ldots,s_m$ are terms with variables (e.g. tuple of terms).

In this setting we use the atomic formula $p(t)$ to represent a process instance with state $p$ and local data $t$.
Furthermore, we use $p_1(t_1),\ldots,p_n(t_n)$ to represent a multiset of atomic formulas.

This kind of transition systems (or extensions of them) have been used to model concurrent processes with local data (identifiers, time-stamp) in models like Timed Networks, Data Nets, MSR(${\mathcal C}$).

In this setting it could be interesting to defined history information that keep tracks of data occurring in the current configuration.
This is what is often needed to verify properties like correspondence in protocol verification, i.e., principals complete protocols maintaining the same nonce, identifier, etc.

For instance,  consider rule
$$
p(x),q(y)\rightarrow p'(x),q'(x)
$$
in which $x,y$ are existentially quantified variables.
This rule can be used to specify a synchronization step in which a process in state $p$ passes its local data 
to a process in state $q$.
To keep track of this event, we add predicates that maintain information about data.
For instance, the rule $p(x),q(y)\rightarrow p'(x),q'(x)[h_{p,q}(x)]$
adds a predicate $h_{p,q}(x)$ to the history keeping track of the data exchanged during the synchronization step. In this setting, when considering conditions that could be used to obtain wsts, we cannot keep state and histories separated.  
In general a rule
$$
p_1(t_1),\ldots,p_n(t_n)\rightarrow q_1(s_1),\ldots,q_m(s_m)[e]
$$
in which $e$ is a predicate that shares variables with $t_1,\ldots,t_n,s_1,\ldots,s_n$,
is translated into the MSR(Id) formula
$$
p_1(t_1),\ldots,p_n(t_n)\rightarrow q_1(s_1),\ldots,q_m(s_m),e: true
$$
When all predicates occurring in the resulting rewriting rules are monadic, then 
HCOV can be decided by resorting the the decision procedures for MSRCOV.
\newcommand{\pgraph}{{\cal G}} 
\newcommand{\pedge}{\rightharpoonup} 

We consider here an example presented in \cite{Delzanno15} that describes how MSR can be applied to
track data in a computation in order to discover or prove absence of permission conflicts in abstract models
of component-based systems (inspired to the Android SO).
We consider a process of type $C$ that handles the contents of a device.
A process of type $I$ represents a potential intruder. We assume here that $C$ and $I$ have incompatible permissions, e.g. $C$ can access the device data whereas $I$ cannot.
If during a computation an identifier  is transferred from a process of type $C$ to a process of type $I$, then the system may behave incorrectly.
In our abstraction of activities, we just need one local data for component used to store received data.
The content component contains an identifier associated to the device private data.
Since each component is defined by send/receive operations only, the MSR(Id) model
consists of the following rewriting rules:
$$
\begin{array}{l}
c_1(x),a_1(y),ok\rightarrow c_1(x),a_2(x),h_{a}(x),ok:true\\
a_2(x),b_1(y),ok\rightarrow a_3(x),b_2(x),h_{b}(x),ok:true\\
b_2(x),i_1(y),ok\rightarrow b_3(x),i_1(x),h_{i}(x),ok:true
\end{array}
$$
where 
$c_1$ is the single state of process type $C$, 
$a_1,a_2,a_3$ are the states of an intermediate process of type $A$ (application) that invokes the services of the content provider $C$, $b_1,b_2,a_3$ are the states of an intermediate process of type $B$ that receive data from the application and sends them over the internet, 
and $i_1$ is the single state of process type $I$ (it represents an intruder or simply access to Internet).

The initial configuration is defined via the following rules:
$$
\begin{array}{l}
init\rightarrow init,max(x):true\\
init,max(x)\rightarrow q(x),max(y):x<y,\ \ q\in\{a_1,b_1,c_1,i_1\}\\
init,max(x)\rightarrow ok:true
\end{array}
$$
These rule assign distinct identifiers to each instance of every type of process.
Starting from $init$ we can generate any number of instances of processes of type $A$, $B$, $C$, and $I$.
The following rule specifies a conflict detection due to information leaking 
from the content provider to the internet-component.
$$
h_c(x),h_i(x)\rightarrow conflict
$$
Checking for possible detection can be done by executing a symbolic backward exploration that 
exploits the constrained multiset $h_c(x),h_i(x):true$ as a symbolic representation of all possible
larger configurations containing instances of $C$.
The computation of predecessors is fully automated. Furthermore, termination is guaranteed by the 
well-structured property of monadic MSR(Id) proved in \cite{DelzannoMSR}.

For the considered example, we perform the following experiments.
First of all, the rewriting rules are represented in Prolog as the following 
set of facts.

\begin{verbatim}
rule([c1(X),a1(_)],[c1(X),a2(X),ha(X)],{},1).
rule([b1(_),a2(X)],[b2(X),a3(X),hb(X)],{},2).
rule([b2(X),i1(_)],[b3(X),i1(X),hi(X)],{},3).
\end{verbatim}

We omit here the initialization phase to simplify the analysis
(e.g. we can omit the \verb+ok+ predicate).
The seed of backward search is the fact 
\verb+f(0, [hc(A), hi(A)], {}, 1, 0, 0)+.
A fact $f(i,m,c,n,r,f)$ denotes a multiset constraint $m:c$ 
computed at step $i$ of the analysis, with order number $n$, 
obtained by applying rule $r$ backwards to a non-deterministically chosen 
submultiset of the multiset constraint contained in fact $f$.
Each fact $f(i,m,c,v_1,v_2,v_3)$ is a representation of an infinite set of configurations obtained
by first taking an instantiation $m_1$ of the formula $m:c$ and then by taking any multiset 
$m'=m_1\oplus m_2$ for any multiset $m_2$.

The symbolic backward engine computes all predecessors in three steps:

\begin{verbatim}
f(3, [c1(A),a1(_),b1(_),i1(_),hc(A)], {}, 4, 3, 1).
f(2, [b1(_),a2(A),i1(_),hc(A)], {}, 3, 2, 2).
f(1, [b2(A),i1(_),hc(A)], {}, 2, 1, 3).
f(0, [hc(A),hi(A)], {}, 1, 0, 0).
\end{verbatim}

The constraint \verb+{}+ is equivalent to $true$.
The symbol \verb+\_+ corresponds to an anonymous free variable.
Initial configurations are contained in the resulting infinite set of configurations.
From the fixpoint, we can build a trace from an initial configuration to a conflict. 
We just have to follow the history of the predecessor computation.
Fact $4$ is generated from fact $3$ via rule $1$.
Fact $3$ is generated from fact $2$ via rule $2$.
Fact $2$ is generated from fact $1$ via rule $3$.
In the trace we can verify that an identifier can move from an instance of a content 
component to an instance of an internet component yielding a violation that cannot 
be detected by using the underlying permission model.

To avoid conflicts, we can modify the definition of the A and B processes
so that the start method is invoked without adding data in the intent.
The resulting rules (in Prolog notations) are as follows.
\begin{verbatim}
rule([c1(X),a1(_),,ok],[c1(X),a2(X),hc(X),ha(X),ok],{},1).
rule([b1(Z),a2(X),ok],[b2(Z),a3(X),ok],{},2).
rule([b2(X),i1(_),hp(X),ok],[p3(X),i1(X),hp(X),hi(X),ok],{},3).
\end{verbatim}
In the second rule instances of A and B synchronize with no data exchange
(each process keeps the old value in its register).
Via the analysis with backward search, we now get the following fixpoint:

{\begin{verbatim}
f(3, [c1(_),a1(_),ok,b1(A),i1(_),hc(A)], {}, 4, 3, 1).
f(2, [b1(A),a2(_),ok,i1(_),hc(A)], {}, 3, 2, 2).
f(1, [b2(A),i1(_),ok,hc(A)], {}, 2, 1, 3).
f(0, [hc(A),hi(A)], {}, 1, 0, 0).
\end{verbatim}}

Fact $3$ has only instances in initial states ($c1$, $a1$, $p1$, $i1$) thus is  candidate to contain 
denotations of initial configurations.
However in fact $3$, $b1$ of type $B$ has an identifier shared with footprint $hc$ associated to type $C$.
By definition, in initial configurations each identifier has the type associated to process in which it is stored.
Thus, no  instance of the pattern represented by fact $3$ can be an initial state.
Namely, any multiset $m\oplus m'$ s.t. $m$ is an instances of 
\verb+[c1(_),a1(_),ok,p1(A),i1(_),hc(A)]+  cannot be an initial state.
The same holds for fact $0$, its denotation cannot contain initial configurations (it is not possible that 
the same identifier belongs to different footprints in an initial configuration).
Since symbolic backward reachability generates all symbolic predecessors of upward closed sets of configurations, the fixpoint is a proof that the modified model is conflict-free  for any number of nodes in initial configurations. 
\subsubsection{Liveness Properties in Parameterized Systems}
Let us go back to Petri Nets-like models in which all processes are indistinguishable black tokens,
or simply a predicate in MSR.
Introducing identifiers in a formulation of their semantics in which the transition systems maintains a log of events can be useful to apply wsts theory to validate properties like responsiveness.
For instance, assume that rewriting rules expressing local transitions are formulated as
$$
p_1(x)\rightarrow p_2(x)
$$
and rules expressing synchronization are expressed as $p_1(y),q_1(x)\rightarrow p_2(x),q_2(y)$.
In this setting we use the atomic formula $p(x)$ to represent a process instance with identifier $x$.
We can now insert events in order to keep track of properties of individual processes.
For instance,  $p_1(x)\rightarrow p_2(x),req(x)$
could be use to record that process $x$ has entered a given section of its code (e.g. request to enter critical section).
A similar rule can be used to mark that the process enters another critical section of its code
$p_1(x)\rightarrow p_2(x),ack(x)$.
We can now apply HCOV to check for the existence of computations in which a process
manages to reach the critical section. The considered target state can be symbolically represented
as the constrained multiset $req(x),ack(x):true$.

\subsubsection{Correspondence Properties}
We now show how to instantiate the approach to model correspondence properties, i.e.,
properties that require a match between two or more actions. 
A typical example in protocol analysis is the requirement that if agent $A$ receives an ack, 
then the receiver has received the message sent by $A$.
Consider as an example a scenario in which two principals, Alice and Bob, want to share a common secret.
We use predicate $a_i$ and $b_i$ to denote states of the two principals.
We abstract away the representation of secrets and keys.
Alice is defined by the following rules:
$$
\begin{array}{l}
a_0,nonce(x) \rightarrow a_1(x),req(x),nonce(x'):x'>x\\
a_1(x),ack(x) \rightarrow a_2(x):true\\
\end{array}
$$
Bob is defined by the following rule:
$$
\begin{array}{l}
b_0,req(x)\rightarrow b_1(x),ack(x):true\\
\end{array}
$$
We can now add events that store complete information about source and destination of messages.
For instance, $h(msg,agent,nonce)$ can be used to denote type, sender and nonces of messages.
$$
\begin{array}{l}
a_0,nonce(x) \rightarrow a_1(x),req(x),nonce(x') [h(req,a,x)]:x'>x\\
a_1(x),ack(x) \rightarrow a_2(x) [h(ack,a,x)]:true\\
b_0,req(x)\rightarrow b_1(x),ack(x) [h(req,b,x),h(ack,b,x)]:true
\end{array}
$$
Assume now that a third type of agents can intercept messages sent by Alice and Bob.
Trudy has the following behavior:
$$
\begin{array}{l}
t_0,req(x)\rightarrow t_1(x),ack(x) [h(req,t,x)]:true
\end{array}
$$
Using the embedding in MSR(Id), we can now check HCOV
to check if there are successful protocol runs in which correspondence is violated,
i.e., is it possible to reach configurations with history that are larger or equal 
to the following one:
$$
a_2(x) [h(req,t,x)]:true
$$
This configuration can be used to show that some of the conversations (identified by the nonce $x$) 
between agents $a$ and $b$ have been intercepted by agent $t$, i.e., Alice succesfully terminate the protocol
but Bob has not received the message.

In the previous example we can reduce the specification to a model with monadic predicates assuming that
principal names and message types range over a finite alphabet. In other words only nonces range over unbounded
set of values and predicates in the history can be rewritten as $h_{msg,ag}(x)$ for $ag,msg$ taken from a finite set.
In this special case we can decide HCOV by using the symbolic backward reachability algorithm for MSR(Id).
\section{Conclusions}
In this paper we studied a new application of wsts to transition systems with history information.
Historical information is used to express properties that relate states generated in different steps of a computation.
States and events can share information. This makes verification more difficult to handle.
To overcome the difficulties, we have shown that it is sometimes possible to deduce positive results by using existing wsts 
as meta-languages for expressing transition systems with events.
Our analysis lies in between wsts with external memory and results obtained when reasoning of sequences of transitions 
in Petri Nets. A peculiarity of our approach is that we consider history information that can depend on elements of 
the current configurations. This can be done to define time-stamps or to handle events that contain data taken from configurations. 
\smallskip\\
{\em Related Work}
The presented paper shares similarities with recent work on parameterized  verification of provenance in distributed applications, history automata and types, and formal models with external memory.
We discuss below these other lines of research.
Parameterized verification of provenance in distributed applications has been considered in \cite{MMW13}.
In this setting regular languages are used as a formal tool to analyze the provenance of messages taken from a finite alphabet. Lifting the idea to parameterized verification yields models  based on Petri Nets in which counters are used
to keep track of state of processes and current step of automata associated to policies. Using regular languages
allows to define complex policies to regulate the flow of messages in a network.
The use of predicates to observe the history of data share similarities with approaches based on history expressions  introduced in \cite{BDFZ09}.  Register Automata and History-Register Automata have also been used to model programs with dynamic allocation in \cite{T11,TG13}. 
Verification of models with external memory has been considered e.g. in \cite{ChadhaV09}.
The external memory is used here to keep track of asynchronous invocations during a program execution (pushdown system).
The main difference with the above mentioned work is that in our setting we restrict the class of properties in order to generalize history information so as to maintain relations defined over data occurring in states and events.
Furthermore, we have formulated conditions that can be used to obtain positive results by combining conditions on 
transition systems and histories.
Our results are obtained via an application of the theory of well-structured transition systems and via 
reductions to low level concurrency models like rewriting systems in which it is possible to manipulate data taken from an infinite ordered domain of identifiers like MSR(Id) \cite{BozPhD,DelzannoMSR,ADV11}. MSR(Id) is also strictly related to $\nu$-nets \cite{Velardo} that provide fresh name generation and equality constraints. The relation between MSR(Id) and $\nu$-nets is studied in \cite{DR13}. 
As shown in \cite{ADV11}, the MSR(Id) model is strictly more expressive than Petri Nets and it has the same expressive power of Datanets \cite{LNORW08}, an extension of Petri Nets with ordered data.

\bibliographystyle{eptcs}
\bibliography{bibliography}

\end{document}